\documentclass[preprint2]{aastex}

\usepackage{apjfonts}

\newcommand{\eboo}{\mbox{$\eta$~Boo}}
\newcommand{\Msol}{\mbox{${M}_\odot$}}
\newcommand{\half}{{\textstyle\frac{1}{2}}}
\newcommand{\sixth}{{\textstyle\frac{1}{6}}}
\newcommand{\tenth}{{\textstyle\frac{1}{10}}}
\newcommand{\Dnu}[1]{\Delta \nu_{#1}}
\newcommand{\dnu}[1]{\delta \nu_{#1}}
\newcommand{\muHz}{\mbox{$\mu$Hz}}
\newcommand{\numax}{\mbox{$\nu_{\rm max}$}}
\newcommand{\nuac}{\mbox{$\nu_{\rm ac}$}}
\newcommand{\acena}{\mbox{$\alpha$~Cen~A}}
\newcommand{\acenb}{\mbox{$\alpha$~Cen~B}}
\newcommand{\aboo}{\mbox{$\alpha$~Boo}}
\newcommand{\bhyi}{\mbox{$\beta$~Hyi}}
\newcommand{\nuind}{\mbox{$\nu$~Ind}}
\newcommand{\xihya}{\mbox{$\xi$~Hya}}
\newcommand{\auma}{\mbox{$\alpha$~UMa}}
\newcommand{\zetahera}{\mbox{$\zeta$~Her~A}}
\newcommand{\deltaeri}{\mbox{$\delta$~Eri}}
\newcommand{\Teff}{\mbox{$T_{\rm eff}$}}

\slugcomment{Invited review, submitted to PASA}

\shorttitle{Solar-like Oscillations}
\shortauthors{Bedding \& Kjeldsen}

\begin{document}

\title{Solar-like Oscillations\\\normalsize (Invited Review to appear in PASA)}

\author{ Timothy R. Bedding\altaffilmark{1} and Hans
Kjeldsen\altaffilmark{2,3} }

\altaffiltext{1}{School of Physics A28, University of Sydney, NSW 2006,
Australia}

\altaffiltext{2}{Teoretisk Astrofysik Center, Danmarks Grundforskningsfond,
8000 Aarhus C, Denmark; {\tt hans@phys.au.dk}}

\altaffiltext{3}{Institut for Fysik og Astronomi, 8000 Aarhus C, Denmark}

\begin{abstract} 
The five-minute oscillations in the Sun have provided a wealth of
information about the solar interior.  After many attempts, positive
detections of similar oscillations in solar-type stars have now been made.
This review discusses the properties of solar-like oscillations, the
methods used to observe them and the results on individual stars.  We
conclude that the study of solar-like oscillations from the ground and
space has an exciting future.
\end{abstract}


\section{Introduction}

Measuring stellar oscillations is a beautiful physics experiment.  A star
is a gaseous sphere and will oscillate in many different modes when
suitably excited.  The frequencies of these oscillations depend on the
density, temperature, gas motion and other properties of the stellar
interior.  The amplitudes of the oscillations are determined by the
excitation and damping processes, which may involve turbulence from
convection, opacity variations and magnetic fields.  Studying the
frequencies and amplitudes of oscillations in different types of stars
promises to lead to significant advances in our understanding of stellar
structure and evolution.

The best targets are stars which oscillate in several modes simultaneously.
The frequency of each mode depends on the mode structure (e.g., the number
of nodes) and on spatial variations of the sound speed and buoyancy
frequency within the star.  The frequency spectrum therefore places strong
constraints on the internal structure.  This analysis, called
asteroseismology, yields information about composition, age, mixing and
internal rotation that cannot be obtained in any other way and is
completely analogous to the seismological study of the interior of the
Earth.  For reviews, see \citet{B+G94} and \citet{G+S95,G+S96},
%and \citet{ChD2003},
and the proceedings of recent conferences
\citep{ABChDLeuven2002,TCMPorto2003}.

\begin{figure*}%[!ht]
\centerline{\includegraphics[width=0.8\textwidth]{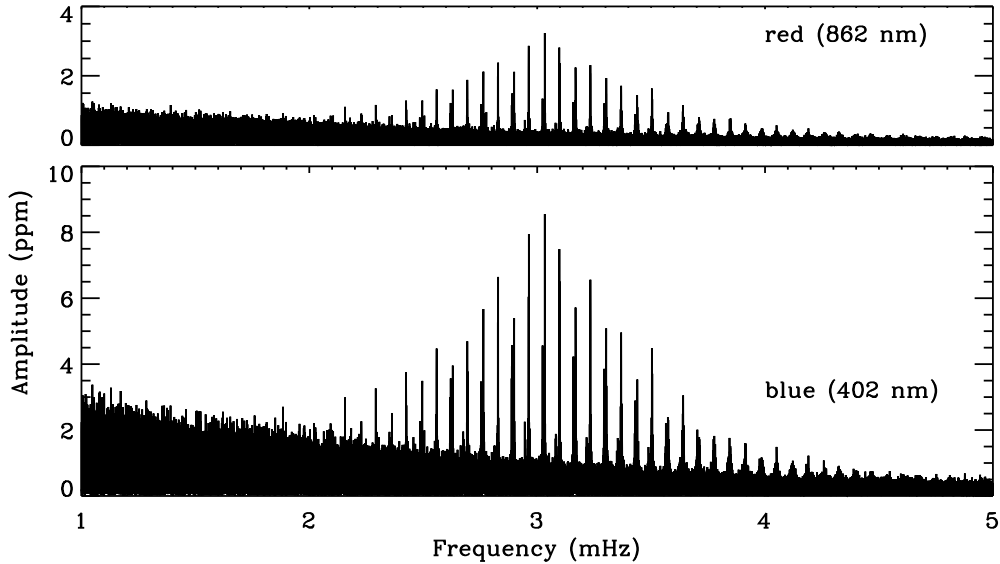}}
\caption[Virgo red and blue spectra]{
\label{fig.VIRGO}
Amplitude spectra of full-disk solar oscillations measured in intensity by
the VIRGO instrument on the SOHO spacecraft.  The observations are smoothed
and rescaled here to show the spectrum corresponding to 30 days.
Individual oscillation modes appear as strong peaks rising above a sloping
background, which arises from random convective motion on the solar
surface.  (Adapted from \citealt{FAA97}.)  }
\end{figure*}

The best-studied example of an oscillating star is the Sun.  It oscillates
in many modes simultaneously, with periods in the range 3--15 minutes.  The
frequencies of the modes contain information about the sound speed and
rotation deep within the Sun, with each mode sensing these quantities in
subtly different ways \citep[e.g.,][]{G+T91,ChD2002,ESAGONG2002}.  Very
accurate measurements of the frequencies have allowed inversion to
determine the internal structure and rotation of the Sun.  Helioseismology
has provided tests of theoretical models of solar evolution and led to
fascinating new insights into the complex dynamics of solar internal
rotation.  Observations that do not resolve the solar disk restrict us to
observing modes of the lowest degree, but these still contain a great deal
of information.  The frequency spectrum of the Sun observed in integrated
light is shown in Fig.~\ref{fig.VIRGO}.

The solar oscillations are excited by convection near the surface, and it
is therefore to be expected that all stars having such a convection zone
will show similar oscillations.  We use the term {\em solar-like
oscillations\/} to refer to oscillations excited stochastically by
convection.  This presumably includes all stars cool enough to have an
outer convection zone, from roughly the cool edge of the classical
instability strip out to the red giants.  However, at least in stars of
roughly solar type, the expected amplitudes are tiny, which places extreme
demands on instrumental stability.  In recent years, however, success has
finally come.  In this review we will describe the main results that have
been achieved so far and make some comments about future prospects.

\section{Properties of oscillations}	\label{sec.properties}

\begin{figure*}%[!ht]
\centerline{\includegraphics[width=0.8\textwidth]{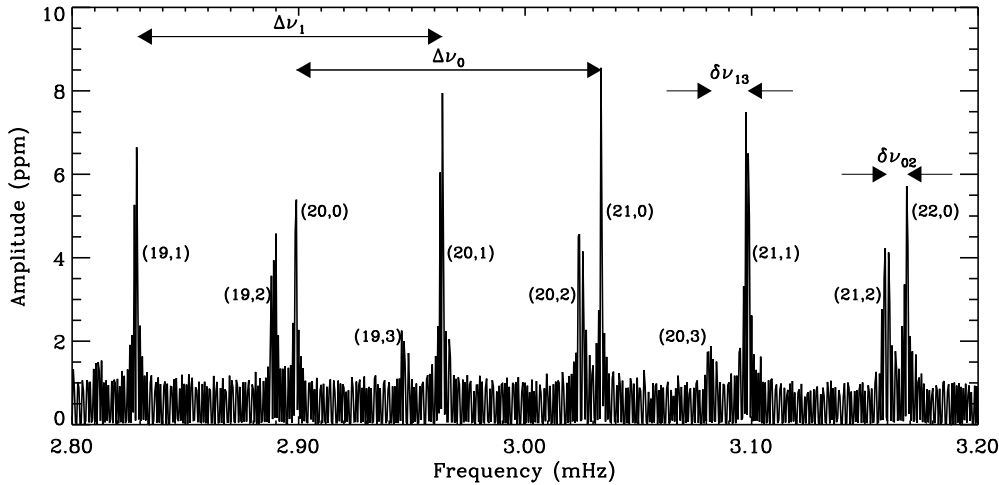}}
\caption{\label{fig.schemspec} Small section of the solar amplitude
spectrum (lower panel of Fig.~\ref{fig.VIRGO}), showing $(n,l)$ values for
each mode.  The large and small separations are indicated.  These measure
the average density and core composition, respectively, and can therefore
be used to infer the mass and age of a star.  }
\end{figure*}

Solar-like oscillations are standing sound waves, known as p~modes (since
pressure is the restoring force).  Each mode is characterised by three
integers: the radial order~$n$, the angular degree~$l$ and the azimuthal
degree~$m$.  Mode frequencies for low-degree p-mode oscillations are
approximated reasonably well by the asymptotic relation:
\begin{equation}
  \nu_{n,l} = \Dnu{} (n + \half l + \epsilon) - l(l+1) D_0.
        \label{eq.asymptotic}
\end{equation}
Here, $\Dnu{}$ (the so-called large separation) reflects the average
stellar density, $D_0$ is sensitive to the sound speed near the core and
$\epsilon$ is sensitive to the surface layers.  It is conventional to
define $\dnu{02}$, the so-called small separation, as the frequency spacing
between adjacent modes with $l=0$ and $l=2$.  These separations are shown
in Fig.~\ref{fig.schemspec}, together with the similar quantity~$\dnu{13}$.
We can further define $\dnu{01}$ to be the amount by which $l=1$ modes are
offset from the midpoint between the $l=0$ modes on either side (not shown
in Fig.~\ref{fig.schemspec}).  If the asymptotic relation holds exactly,
then it follows that $D_0 = \sixth\dnu{02} = \half\dnu{01} =
\tenth\dnu{13}$.

In practice, the asymptotic relation does not hold exactly, even for the
Sun.  For example, the large separation depends on~$l$ (hence the separate
definitions of $\Dnu{0}$ and $\Dnu{1}$ in Fig.~\ref{fig.schemspec}).  The
large and small separations also depend on frequency.
%, which leads to curvature in the echelle diagram.  
Therefore, a first asteroseismic analysis might involve comparing $\Dnu{}$
and $\dnu{02}$ with models \citep[e.g.,][]{BChDW94,Gou2003}, but a full analysis
will include a detailed comparison of individual frequencies.

%\citet{R+RC2002}

\begin{figure}[tb!]
\includegraphics[bb=69 65 577 712,angle=90,width=\the\hsize]{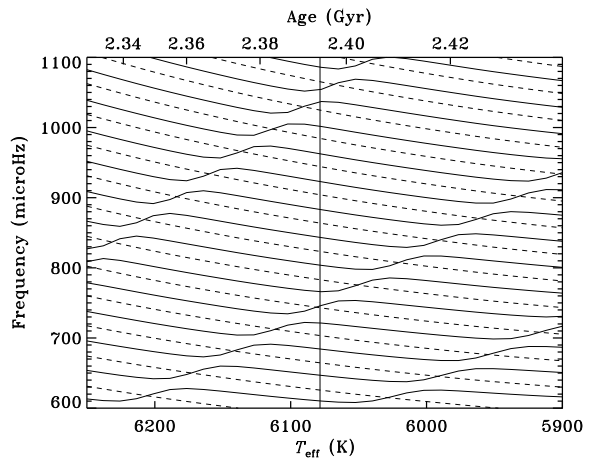}
\caption[]{\label{fig-crossings} Evolution of adiabatic frequencies with
  age of a model of mass $1.60 \Msol$, where age is measured by the
  effective temperature.  The dashed lines correspond to modes of degree $l
  = 0$, and the solid lines to $l = 1$.  The vertical solid line indicates
  the $\Teff$ of \eboo.  Figure taken from \citet{ChDBK95}. }
\end{figure}

Significant departures from the asymptotic relation are expected for
evolved stars.  As discussed by \citet{ChDBK95}, the steep maximum in the
local buoyancy frequency at the edge of the inert helium core gives rise to
modes behaving like trapped internal gravity waves (so-called g~modes;
e.g., \citealt{D+P91,APChD95}).  The core contracts as the star evolves,
leading to an increase in the buoyancy frequency and hence in the
frequencies of the g~modes.  Meanwhile, the p-mode frequencies decrease
with time as the mean density decreases.  When the frequency of a g~mode
approaches that of a p~mode, the two modes undergo an avoided crossing
(also called mode bumping) where they exchange physical nature.  This is
illustrated in Figure~\ref{fig-crossings}, which shows frequency as a
function of $\Teff$ for an evolving $1.6\Msol$ star.  Radial modes ($l =
0$), shown by dashed lines, closely follow equation~(\ref{eq.asymptotic});
the decrease in frequency with age reflects the increasing stellar radius.
The same trend is seen for modes with $l = 1$ (solid lines), but here we
also see the effect of the additional set of g~modes, whose frequencies
increase with age.  At the avoided crossings the modes take on mixed
characters between pure~p and pure g~modes.  The frequencies of these mixed
modes are extremely sensitive to the evolutionary state and are therefore
very interesting, provided they are excited to observable amplitudes.
Observational evidence for mixed modes, in the form of frequencies
departing from the asymptotic relation, has been seen in subgiants such as
\eboo{} and \bhyi.

We also note that stellar rotation causes modes with $l \ge 1$ to split
into multiplets with different values of~$m$, with a separation that
directly measures the rotation rate averaged over the region of the star
that is sampled by the mode.  For the Sun, these types of measurements have
given important information about the internal rotation
\citep[e.g.,][]{CEI2001} and it is hoped that similar results will soon be
achieved on other stars \citep[see][]{G+S2003}.  The measurements are
particularly difficult because they require long time series, in order to
resolve the rotational splittings.

Oscillations in the Sun are long-lived compared to their periods, which
allows their frequencies to be measured very precisely.  However, the
lifetime is finite and this results in the peaks in the power spectrum
having a Lorentzian profile with a linewidth that indicates the mode
lifetime \citep{T+F92}.  Mode lifetimes have not yet been directly measured
for other solar-type stars, and a concern remains that some stars may have
mode lifetimes so short that rotational splittings -- and perhaps even the
small and large separations -- will not be observable.  In the case of red
giants, on the other hand, it appears that mode lifetimes have been
measured (see Sec.~\ref{sec.semireg}).

\section{Observational techniques}

Stellar oscillations have been observed by their effect on the stellar
atmosphere in three ways \citep[see e.g.,][]{Kje2003}: (i)~velocity shifts
of spectral lines, (ii)~variations in total intensity (photometry) and
(iii)~variations in equivalent widths of temperature-sensitive lines.  In
all cases, coverage of the target needs to be as continuous as possible to
reduce confusion from aliases in the spectral window.\footnote{The times at
which observations are made define the {\em window function}, and the
transform of this in the frequency domain is the {\em spectral window}.}

\subsection{Velocity measurements}

Most of the new results are based on velocity measurements obtained using
high-dispersion spectrographs with stable reference sources.  The dramatic
improvement in Doppler precision over recent years is a direct result of
programs to detect planets around other stars.  Asteroseismology has
benefited tremendously from these advances.  Indeed, in one sense it is
easier to measure oscillations because the timescales of interest are much
shorter than for exoplanets.  On the other hand, the amplitudes of
oscillations in solar-type stars are extremely small, which is the reason
that positive detections have been so difficult.

\subsection{Intensity measurements}

Intensity measurements have three tremendous advantages: they can be made
with extremely simple instrumentation; they use photons across a wide range
of wavelengths; and they can be made simultaneously on many stars (although
to some extent, multi-object spectroscopy also offers this possibility).
Unfortunately, scintillation from the Earth's atmosphere severely limits
the precision of ground-based photometry.  The best results to date were
obtained by \citet{GBK93} in a multi-site campaign using differential
photometry of the open cluster~M67, but these fell short of providing
definite detections.  Atmospheric scintillation on the Antarctic plateau
may be small enough to make high-precision photometry feasible
\citep{Mar2002}, and there are plans to establish an observatory at Dome~C
\citep{SFV2002}.  The chance for many weeks -- or even months -- of
continuous observations represent an additional bonus.

Space is the ideal place to make intensity measurements, given the absence
of atmospheric scintillation and the possibility for long periods of
uninterrupted observations.  The Fine Guidance Sensors on the Hubble Space
Telescope were exploited by \citet{ZKW99} to perform high-precision
photometry, as was the 52-mm star camera on the failed {\em WIRE\/}
satellite \citep{Buz2002}.  The space missions designed specifically for
asteroseismology are {\em MOST\/} \citep[][launch date June 2003]{MKW2000},
{\em COROT\/} \citep{Bag98} and {\em Eddington\/} \citep{R+F2003}.  It is
unfortunate that {\em MONS\/} \citep{KBChD2000} appears unlikely to
proceed, due to lack of funding.

\subsection{Equivalent-widths}

This method for detecting oscillations was suggested by \citet[][see also
\citealt{Ste2002}]{KBV95}.  It involves monitoring changes in spectral
lines whose equivalent widths (EWs) are temperature-sensitive, most notably
the hydrogen Balmer lines.  Observations of \eboo{} showed good evidence
for oscillations (see Sec.~\ref{sec.eboo}), while those of \acena{} only
produced a tentative detection \citep{KBF99}.

Meanwhile, the usefulness of the Balmer lines for measuring oscillations
was questioned by \citet{HJLaB96}.  They cited the study by \citet{RHD91}
of the wavelength variation of oscillations in the Sun, which showed that
the Balmer lines remain stable while the surrounding continuum fluctuates.
Indeed, this is precisely the effect that makes Balmer-line EWs so useful
\citep[see also][]{T+L2002}.

\subsection{Comparison of the methods}

Velocity and EW measurements have an important advantage: they are both
more sensitive to modes with $l \ge 2$ than are intensity measurements.  In
both cases, the reason is that the observations have some spatial
resolution of the stellar disk, and so the tendency for high-degree modes
to cancel is reduced.  For velocity measurements, we measure velocities
projected onto the line of sight, which gives more sensitivity to the
centre of the disk relative to the limb \citep{CHd+G82}.  For EW
measurements, at least for the Balmer lines, the centre-to-limb variation
of line strengths gives a similar effect \citep{BKR96}.  As pointed out by
these last authors, the different sensitivities to~$l$ of the various
methods provides a useful tool for mode identification, and this has been
applied to $\delta$~Scuti stars \citep{VKB98,DFL2002,D+F2002}.

Intensity and EW measurements both detect temperature changes, and so are
much more sensitive to the stellar background that arises from granulation.
This is seen in intensity observations of the Sun (Fig.~\ref{fig.VIRGO}) as
background power rising towards low frequencies, and has been detected in
EW observations of \acena{} \citep{KBF99}.  Although this power contains
information about stellar convection that might usefully be compared with
models \citep[e.g.,][]{TChDN98}, for asteroseismology it represents an
unwanted and fundamental noise source.  Velocity observations are far less
sensitive to the stellar background, which is a very important advantage.

Finally, we note one disadvantage of velocity measurements over the other
methods: the precision is degraded by stellar rotation, due to line
broadening.  Of the stars discussed here, only \eboo{} is rotating rapidly
enough for this effect to be significant.

At this stage, it seems clear that measuring velocities is the method of
choice for ground-based observations.  However, the observations require
high-resolution spectroscopy with extremely high precision, which is only
achievable on a handful of instruments.  EW measurements may have a place
in observing rapid rotators, for which velocity measurements become less
precise.  Intensity measurements are planned for space missions, and also
for observations from Antarctica.  Of course, photometry will also continue
to be the main method of observing high-amplitude oscillators such as
semiregulars (Sec.~\ref{sec.semireg}).

\section{Results for individual stars}

\begin{figure}[tb!]
\includegraphics[bb=34 11 431 431,width=\the\hsize] {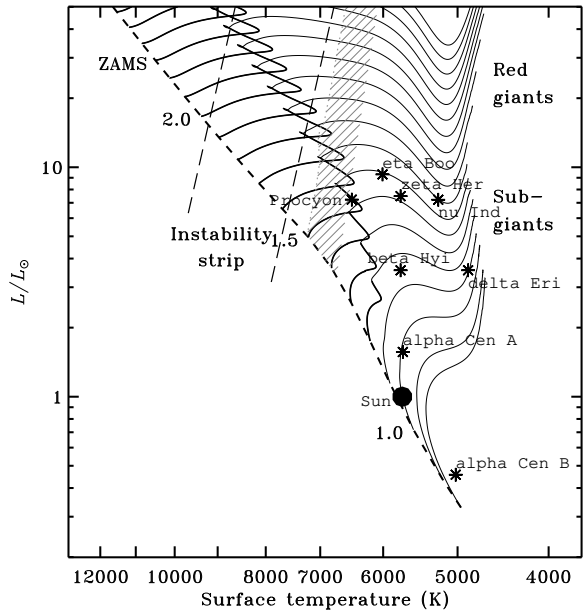}

\caption[]{\label{fig.hr} H-R diagram showing some of the stars discussed
in the text.}
\end{figure}

Early attempts to measure solar-like oscillations, including some claims
for detections, have been reviewed by \citet{B+G94} and \citet{K+B95}.
More recent reviews appear in conference proceedings and these can be used
to follow the development of the field
\citep{Fra97,K+B97,Bro98,B+K98,Bro2000,K+B2000,B+C2003}.  Here we discuss
results on some individual stars, with emphasis on the most recent results.
The positions in the H-R diagram of most of these stars are shown in
Fig.~\ref{fig.hr}.  Most of the observations discussed here were made in
velocity.  The exceptions are the EW observations of \eboo{} and the
photometric observations of \acena{} and of the red giants (\aboo, \auma{}
and the semiregulars).

This review deals mostly with observational aspects, although papers
describing theoretical models of individual stars are noted in the relevant
sections.  There are several reviews in conference proceedings of the
theoretical aspects of asteroseismology of solar-like stars
\citep[e.g.,][]{ChD98a,Tho2000,Rox2002}.  In addition, we refer the reader
to the comprehensive modelling of oscillation frequencies for
intermediate-mass stars by \citet{Gue2002}.

%%%%%%%%%%%%%%%%%%%%%%%%%%%%%%%%%%%%%%%%%%%%%%%%%%%%%%%%%%%%%%%%%%%%%%
\subsection{Procyon ($\alpha$~CMi)}

Procyon is an F5 subgiant and the second brightest star of near-solar type
(the brightest is $\alpha$~Cen).  Evidence for oscillations in Procyon was
presented by \citet{BGN91}, in the form of a broad envelope of excess power
in the Fourier spectrum.  We have previously argued \citep{K+B95} that this
power excess could be due to noise, either from instrumental drift or from
stellar background.  Since then, additional measurements with the FOE and
AFOE spectrographs have continued to show the power excess \citep{Bro2000},
although without revealing the characteristic regular series of peaks.
Meanwhile, measurements with the ELODIE spectrograph by \citet[][see also
\citealp{BMM99}]{MSL99}, gave clear confirmation of the power excess and
also provided evidence for a regular spacing of $\Dnu{} = 55$\,\muHz.  The
power excess has also been seen using the EMILIE and CORALIE spectrographs
\citep{BSB2002,CBK2002} and it now seems clear that it has a stellar
origin, although the signature of p-mode oscillations has not been
confirmed.  It remains to be seen whether the difficulty in obtaining a
power spectrum of Procyon that shows a clear p-mode structure is due to
insufficient high-quality data (i.e., confusion from aliases in the
spectral window) or to intrinsic properties of the star (i.e., short mode
lifetimes).  Finally, we note that several recent papers present
theoretical models for Procyon which are consistent with the
above-mentioned value of $\Dnu{}$ \citep{CDG99,DiM+ChD2001,PMB2002}.

%%%%%%%%%%%%%%%%%%%%%%%%%%%%%%%%%%%%%%%%%%%%%%%%%%%%%%%%%%%%%%%%%%%%%%
\subsection{\eboo}	\label{sec.eboo}

\citet{KBV95} reported evidence for oscillations in the G0 sub-giant
$\eta$~Boo, based on measurements of Balmer-line EWs.  
%
%They presented this as the first clear evidence of solar-like oscillations
%in a star other than the Sun.  The observations were obtained over six
%nights with the 2.5\,m Nordic Optical Telescope on La Palma, and consisted
%of 12684 low-dispersion spectra.  In the power spectrum of the
%equivalent-width measurements, they found an excess of power at frequencies
%around 850\,\muHz.  The average amplitude inferred for the oscillations was
%about 7 times greater than solar, in rough agreement with the empirical
%scaling relation suggested by \citet{K+B95}.  Comb analysis of the power
%spectrum suggested a regular spacing of $\Dnu{}=40.3$\,\muHz.  Based on
%this, they identified thirteen oscillation modes.  Similar observations of
%the daytime sky showed the five-minute solar oscillations at the expected
%frequencies.
%
The observed frequencies, taken with available estimates of the stellar
parameters, were in good agreement with theoretical models
\citep{ChDBK95,G+D96}.  Particularly interesting was the occurrence in
theoretical models -- and apparently in the observations -- of avoided
crossings, in which mode frequencies were shifted from their usual regular
spacing by effects of gravity modes in the stellar core (see
Sec.~\ref{sec.properties}).  
%
%Since then, the improved luminosity estimate for \eboo{} from Hipparcos
%measurements has given even better agreement with the measured value of
%$\Dnu{}$ \citep{BKChD98}.
%
Meanwhile, a search for velocity oscillations in \eboo{} with the AFOE
spectrograph by \citet{BKK97} failed to detect a signal, setting limits at
a level below the value expected on the basis of the EW results.

%Although the data were sparse (22 hours spread over 7 successive nights)
%and the precision was degraded by the relatively fast rotation of the star
%($v\sin i = 13$\,km\,s$^{-1}$), the analysis by \citet{BKK97} was careful
%and thorough, and the results seemed to be inconsistent with those of
%\citeauthor{KBV95}

\begin{figure}[tb!]
\includegraphics[width=\the\hsize, bb=68 369 545 702]{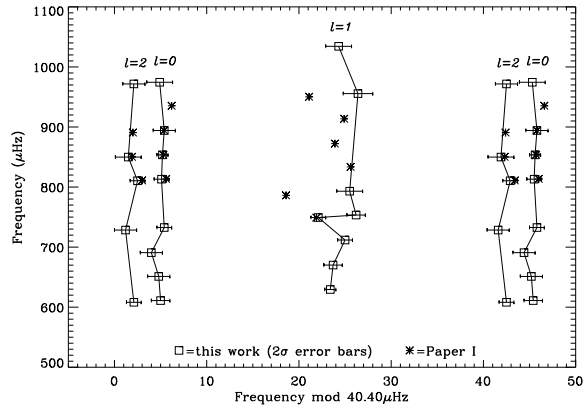}
\caption[]{\label{fig.eboo-echelle} So-called echelle diagram showing the
21 frequencies in \eboo, with $2\sigma$ error bars, together with the 13
frequencies reported by \citet{KBV95}.  Figure from \citet{KBB2003}, and
Paper~I refers to \citet{KBV95}. }
\end{figure}

\citet{KBB2003} presented further observations of \eboo, obtained in 1998 in
both EW and velocity.  They recorded spectra with the Nordic Optical
Telescope, which they used to measure equivalent widths of strong
temperature-sensitive lines.  They also measured velocities using an iodine
reference cell at Lick Observatory.  Their analysis also included velocity
measurements published by \citet{BKK97} and the original equivalent-width
measurements by \citet{KBV95}.  All four data sets showed power excesses
consistent with oscillations, although with a range of amplitudes that
presumably reflects the stochastic nature of the excitation.  The highest
peaks showed regularity with a large separation of $\Dnu{}=40.4$\,\muHz{}
and they identified 21 oscillation frequencies from the combined data (see
Fig.~\ref{fig.eboo-echelle}).  The observations indicate that peak
oscillation amplitudes in \eboo{} are typically 3--5 times solar.  This
conclusion is consistent with the upper limits reported by \citet{BKK97}.

Finally, we note that \citet[][and paper in preparation]{CBE2003} reported
velocity measurements using the CORALIE and ELODIE spectrographs that
showed a clear excess of power and a frequency spacing of $39.6$\,\muHz{}.
We can conclude that oscillations in \eboo{} have been clearly detected and
that individual mode frequencies have been identified.  The frequencies
have been compared with theory by \citet{DiMChDK2003}, who were able to
reproduce them using models both with and without convective-core
overshooting.  Future observations of \eboo, particularly with the {\em
MOST\/} spacecraft, should measure more oscillation modes with greater
frequency precision and permit discrimination between these alternatives.

%%%%%%%%%%%%%%%%%%%%%%%%%%%%%%%%%%%%%%%%%%%%%%%%%%%%%%%%%%%%%%%%%%%%%%
\subsection{$\beta$ Hyi}

A clear power excess in this southern G2 subgiant was detected using
velocity measurements with UCLES at the Anglo-Australian Telescope by
\citet{BBK2001} and confirmed with CORALIE by \citet{CBK2001}.  Although
the amplitudes of the highest peaks were in agreement with theoretical
expectations, \citet{Gou2001} has since pointed out that the total observed
power was a factor of about 1.9 below expectations.  In another
development, \citet{BKB2002} found evidence in the UCLES data for a
short-lived, high-amplitude oscillation event.  Such events have not been
seen in the Sun and, if confirmed as a feature of subgiants, these
`starquakes' would make it harder to measure accurate mode frequencies and
perform asteroseismology.

The UCLES and CORALIE datasets are now being combined to produce a set of
oscillation frequencies (Kjeldsen et al., in preparation).  Meanwhile,
theoretical calculations of frequencies for \bhyi{} have been carried out
by \citet{F+M2003} and by \citet{DiMChDP2003}, both of which indicate the
occurrence of avoided crossings for modes with $l=1$.

Finally, we note that a long-term trend in radial velocity, possibly
indicating a low-mass companion, has been reported by \citet{EKE2002} but
not yet confirmed by other observers \citep[e.g.,][]{JBM2002}.

%%%%%%%%%%%%%%%%%%%%%%%%%%%%%%%%%%%%%%%%%%%%%%%%%%%%%%%%%%%%%%%%%%%%%%
\subsection{\zetahera} 

This G0 subgiant was observed in velocity with the ELODIE and CORALIE
spectrographs \citep{MBP2001,BSL2003}.  Excess power was seen with a
possible large separation, but more observations are needed to establish
the mode frequencies.

%%%%%%%%%%%%%%%%%%%%%%%%%%%%%%%%%%%%%%%%%%%%%%%%%%%%%%%%%%%%%%%%%%%%%%
\subsection{\deltaeri} 

This K0 subgiant was observed in velocity with the CORALIE spectrograph by
\citep{Cetal2003}.  Again, excess power was seen with a possible large
separation, but more observations are needed to establish the mode
frequencies.

%%%%%%%%%%%%%%%%%%%%%%%%%%%%%%%%%%%%%%%%%%%%%%%%%%%%%%%%%%%%%%%%%%%%%%
\subsection{\acena}

\begin{figure}[bt!]
\includegraphics[width=\the\hsize, bb=22 164 562 623]{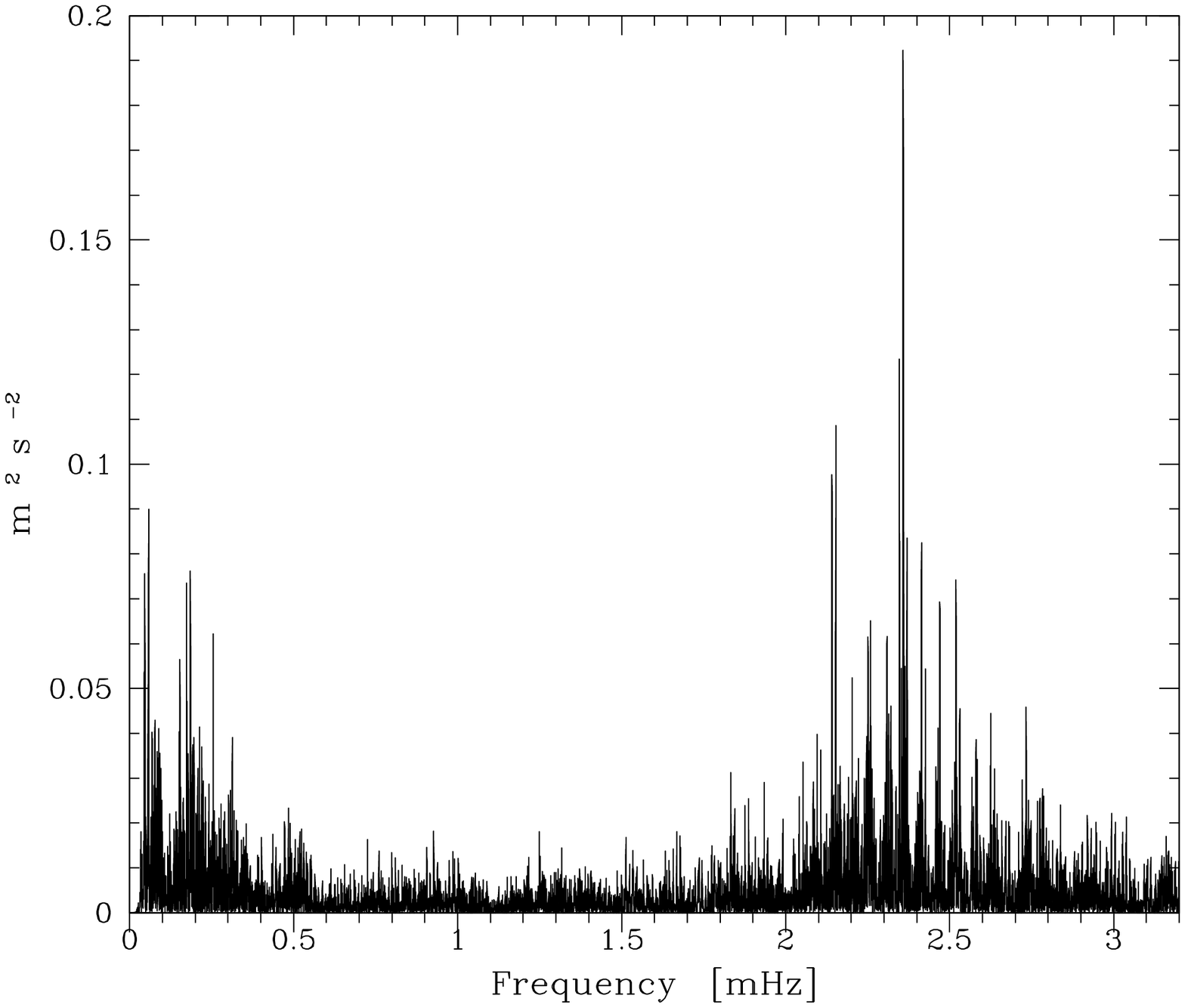}
\caption[]{\label{fig.acena} Power spectrum of \acena, from  \citet{B+C2002}.}
\end{figure}

As a nearby star with the same spectral type as the Sun, \acena{} is an
obvious target for asteroseismology.  The clear detection of p-mode
oscillations by \citet{B+C2001,B+C2002} using velocity measurements with
the CORALIE spectrograph was an important breakthrough (see
Fig.~\ref{fig.acena}).  These results confirmed the earlier but less secure
detection by \citet{S+B2000}, who used photometry from the {\em WIRE\/}
satellite.  The CORALIE data have produced frequencies for 28 modes with $l
= 0$--$2$ and $n=15$--$25$ \citep{B+C2002} which have been compared with
theoretical models by \citet{TPM2002} and \citet{TSN2003}.  Our own group
is analysing additional data on this star taken with the UCLES and UVES
spectrographs, and we expect to increase the number of detected
frequencies.  It is safe to say that seismology of \acena{} will be an
important area of research for many years.

%%%%%%%%%%%%%%%%%%%%%%%%%%%%%%%%%%%%%%%%%%%%%%%%%%%%%%%%%%%%%%%%%%%%%%

\subsection{\nuind}

This metal-poor G subgiant ($\mbox{[Fe/H]} = -1.4$) was suggested as a
target for asteroseismology by \citet{Nis98}.  Oscillations were detected
in velocity using the UCLES and CORALIE spectrographs, and the results are
being analysed (Bedding et al., in prep.; Carrier et al., in prep.).

%%%%%%%%%%%%%%%%%%%%%%%%%%%%%%%%%%%%%%%%%%%%%%%%%%%%%%%%%%%%%%%%%%%%%%
\subsection{\xihya}

This G7~III giant was observed in velocity with the CORALIE spectrograph
\citep{FCA2002,Ste2002}.  The observations showed multi-periodic
oscillations consistent with radial ($l=0$) oscillations \citep[see
also][]{TChDC2003}.  Using a stochastic excitation model, \citet{H+G2002}
found fair agreement between estimated velocity amplitudes and the
observations.  More work is still to be done, but it seems clear that
p-mode oscillations have been detected in this star.

%%%%%%%%%%%%%%%%%%%%%%%%%%%%%%%%%%%%%%%%%%%%%%%%%%%%%%%%%%%%%%%%%%%%%%
\subsection{Arcturus (\aboo)}

\begin{figure}[bt!]
\includegraphics[width=\the\hsize, bb=78 369 545 702]{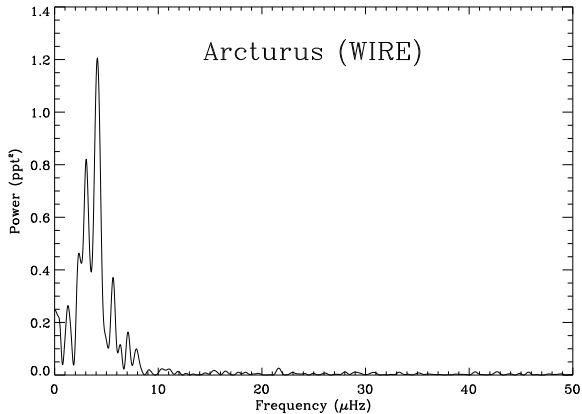}
\caption[]{\label{fig.aboo} Power spectrum of Arcturus, from {\em WIRE\/}
photometry \citep{RBB2003}}
\end{figure}

This bright K1~III giant is known to be variable in radial velocity on
timescales of a few days \citep{BJP90,H+C94a,Mer95}, but without good
evidence for a regular p-mode structure in the power spectrum.
Observations with the star camera on the {\em WIRE\/} satellite show
photometric variability that is consistent with solar-like oscillations
\citep{RBB2003}, and the power spectrum appears to have a regular series of
peaks with a spacing of 0.83\,\muHz{} (Fig.~\ref{fig.aboo}).  However, as
pointed out by \citeauthor{RBB2003}, simulations with pure noise show that
the frequencies extracted from a hump of excess power will tend to be
regularly spaced, with a separation slightly greater than the formal
frequency resolution of the data set (which was 0.64\,\muHz{} in this
case).  Thus, one must be careful in interpreting such a signature, since
the envelope of power in Arcturus could also arise from a single mode with
a short lifetime.  A longer time series is needed to decide this issue.

\begin{figure*}[tb!]
\centerline{\includegraphics[width=0.7\textwidth, bb=65 549 553
716]{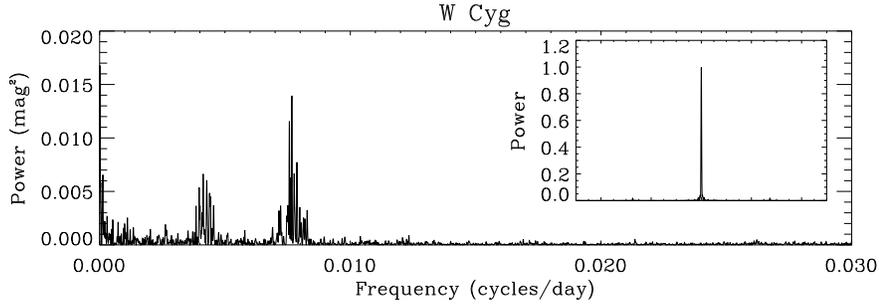}}
\caption[]{\label{fig.sr} Power spectrum of visual observations of the
semiregular variable W~Cyg.  The inset shows the spectral window.  For more
examples, see \citet{Bed2003}.}
\end{figure*}

%%%%%%%%%%%%%%%%%%%%%%%%%%%%%%%%%%%%%%%%%%%%%%%%%%%%%%%%%%%%%%%%%%%%%%
\subsection{\auma}

This K0~III giant was observed in photometry with the star camera on the
{\em WIRE\/} spacecraft by \citet{BCL2000}, who interpreted the variability
as being due to p-mode oscillations.  They reported the apparent detection
of ten oscillation modes with a mean separation of 2.94\,\muHz.  However,
as pointed out in the previous section, care must be taken in interpreting
such a signature.

Little work has been done on theoretical models for red giants.  In
response to the {\em WIRE\/} results, models for \auma{} were calculated by
two groups, but with inconsistent results.  \citet{GDB2000} were able to
match most of the observed frequencies with radial modes, while
\citet{DGH2001} were unable to explain the observed oscillation properties.

%%%%%%%%%%%%%%%%%%%%%%%%%%%%%%%%%%%%%%%%%%%%%%%%%%%%%%%%%%%%%%%%%%%%%%
\subsection{Semiregular variables} \label{sec.semireg}

Oscillating red giants with high luminosity -- the long period variables --
are conventionally divided into Miras and semiregulars.  Mira variables
have large amplitudes and are very regular, reflecting the nature of the
driving process, which is self-excitation via opacity variations.
Semiregulars, on the other hand, have lower amplitudes, less regularity and
often show two or three periods.  In these stars, it seems plausible that
there is a substantial contribution from convection to the excitation and
damping.  Indeed, \citet{ChDKM2001} have suggested that the amplitude
variability seen in semiregulars is consistent with the pulsations being
solar-like, i.e., stochastically excited by convection.  The subject of
Mira-like versus solar-like excitation has also been discussed in the
context of K~giants by \citet{DGH2001}.

Power spectra of the light curves of semiregular variables, based on visual
magnitude estimates spanning many decades, show clear evidence for
stochastic excitation with mode lifetimes ranging from years to decades
\citep{Bed2003}.  An example is shown in Fig.~\ref{fig.sr}.  We expect more
results on semiregulars to be found from mining of databases of visual and
CCD photometry.

\section{Conclusions}

Now that solar-like oscillations have been detected in several stars, we
are in a position to make some definitive statements.  The first and most
obvious is that solar-like stars really do oscillate, with properties
approximately as expected.

Concerning the oscillation amplitudes, recent theoretical calculations have
been made by \citet{HBChD99,Hou2002,H+G2002} and \citet{SGL2003,SNS2003}.
The agreement with observations is fairly good, although it still appears
that F-type stars (Procyon and stars in M67) oscillate with lower
amplitudes than predicted \citep{K+B95,S+H2000}.  Also, oscillations have
yet to be measured in main-sequence stars cooler than the Sun, such as
\acenb.

\begin{figure}[bt!]
\includegraphics[width=\the\hsize,bb= 5 329 480 787]{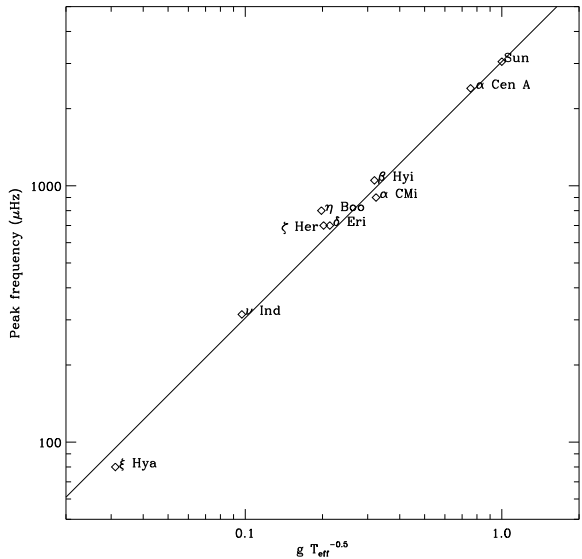}
\caption[]{\label{fig.numax} Observed versus expected peak frequencies,
where expected values are based on scaling the acoustic cutoff frequency.
The diagonal line has a slope of one and passes through the solar value.}
\end{figure}

\begin{figure*}[t!]
\includegraphics[width=0.5\textwidth,bb=122 95 499 747]{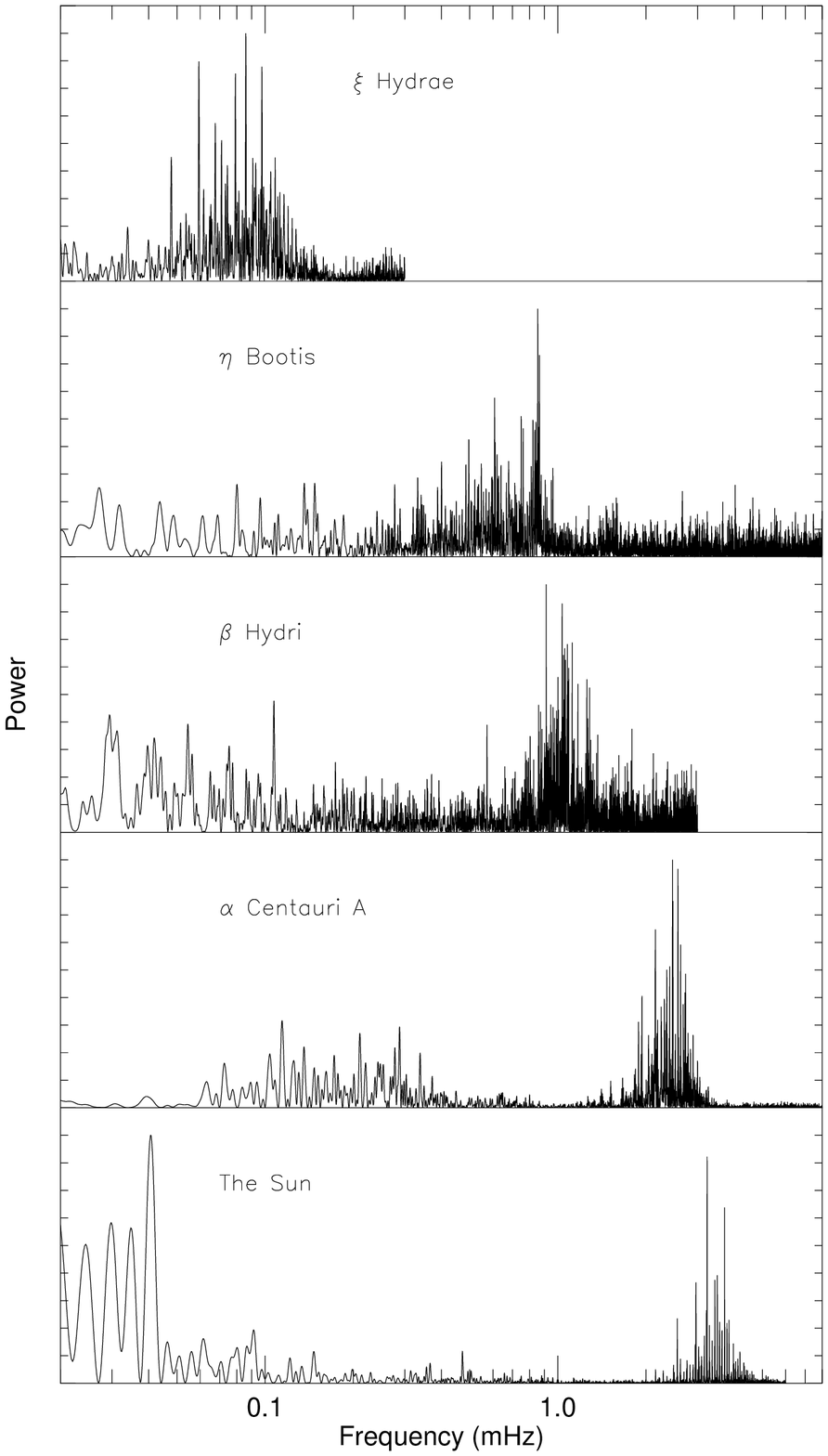} \hfill
\includegraphics[width=0.5\textwidth,bb=122 95 499 747]{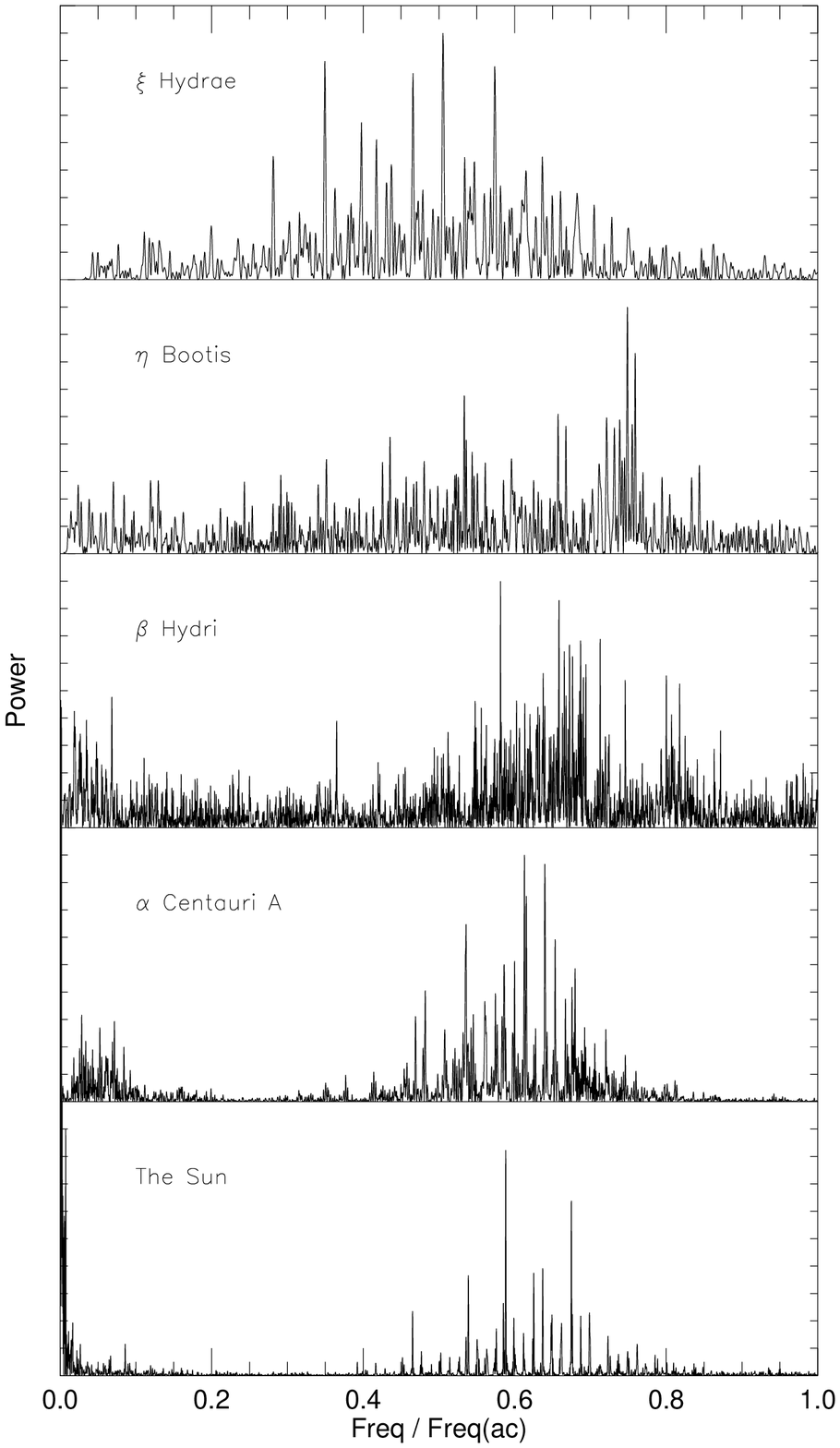}
\caption[]{\label{fig.lots} Observed power spectra of oscillations in the
Sun and four other stars.  On the left, the horizontal axis is observed
frequency on a logarithmic scale.  On the right, we show frequency (on a
linear scale) as a fraction of the acoustic cutoff frequency.  The details
of the data are as follows: Sun -- full-disk velocities from the GOLF
instrument on SOHO (\citealt{GCG97}; note the much lower background
compared to the intensity observations in Fig.~\ref{fig.VIRGO}); \acena{}
-- velocities from UVES and UCLES (Butler et al., in prep.); \bhyi{} --
velocities from UCLES and CORALIE (Kjeldsen et al., in prep.); \eboo{} --
EW from the NOT \citep{KBV95}; \xihya{} -- velocities from CORALIE
\citep{FCA2002}.  Note that the vertical scales have been normalized to
make all stars similar.  The actual amplitudes increase up the plot, and
the peak power in \xihya{} is about 60 times that of the Sun. }
\end{figure*}

The observed frequency of maximum power, on the other hand, is in good
agreement with theory.  The power spectrum of oscillations in the Sun is
modulated by a broad envelope whose maximum is at a frequency of
$\numax\simeq3$\,mHz.  The shape of the envelope and the value of $\numax$
are determined by the excitation and damping.  \citet{BGN91} suggested that
$\numax$ in other stars should scale with the acoustic cutoff
frequency,~$\nuac$.  That is, we expect $\numax \propto g \Teff^{-0.5}$
\citep[see also][]{K+B95}.  As shown in Fig.~\ref{fig.numax}, the observed
oscillations agree well with this scaling relation.  Figure~\ref{fig.lots}
shows the observed oscillation spectra of the Sun and four other stars.

Another result from observations is that avoided crossings in evolved stars
seem to occur and offer excellent prospects for detailed analysis.
However, they complicate the interpretation of the power spectra and demand
observations with few or no gaps, to avoid confusion by aliases in the
spectral window.

Further progress from ground-based observations will come from coordinated
multi-site campaigns on carefully chosen targets \citep[see][]{Pij2003}.
We also look forward with great excitement to the dedicated space missions.
Many years of effort are now paying off, and we foresee a great future for
the study of solar-like oscillations.

%\citet{App2003}

\acknowledgments

This work was supported financially by the Australian Research Council, the
Danish Natural Science Research Council and the Danish National Research
Foundation through its establishment of the Theoretical Astrophysics
Center.  We thank Dennis Stello for helpful comments on the manuscript.

\end{document}